\documentclass[aps,prl,twocolumn,superscriptaddress,showpacs,nofootonbib]{revtex4}
\usepackage{graphicx}

\newcommand{\ket}[1]{\ensuremath{|{#1}\rangle}}

\newcommand{\braket}[1]{\ensuremath{\langle{#1}\rangle}}

\begin{document}

\title{How much entanglement can be generated between two atoms by detecting photons?}

\author{L. Lamata}
\email[]{lamata@imaff.cfmac.csic.es} \affiliation{Instituto de
Matem\'{a}ticas y F\'{\i}sica Fundamental, CSIC, Serrano 113-bis,
28006 Madrid, Spain}\affiliation{Max-Planck-Institut f\"ur
Quantenoptik, Hans-Kopfermann-Strasse 1, 85748 Garching, Germany}

\author{J.J. Garc\'{\i}a-Ripoll}
\affiliation{Max-Planck-Institut
 f\"ur Quantenoptik, Hans-Kopfermann-Strasse 1, 85748 Garching,
  Germany}

\author{J.I. Cirac}
\affiliation{Max-Planck-Institut f\"ur Quantenoptik,
Hans-Kopfermann-Strasse 1, 85748 Garching, Germany}

%\date{\today}

\begin{abstract}
  It is possible to achieve an arbitrary amount of entanglement
  between two atoms using only spontaneously emitted photons, linear
  optics, single photon sources and projective measurements. This is
  in contrast to all current experimental proposals for entangling two
  atoms, which are fundamentally restricted to one entanglement bit
  or ``ebit''.
\end{abstract}

\pacs{03.67.Mn,32.80.Pj,42.50.Ct}

\maketitle

In the world of quantum information processing it is widely accepted
that, while photons are the ideal candidates for transmitting quantum
information, this information is better stored and manipulated using
atomic systems. The reason is that, while photons can be moved through
long distances with little decoherence, atoms can be easily confined
and can preserve quantum information for a long time.  Consequently,
an ideal design for a quantum network will conceivably be built upon a
number of atomic or solid state devices which communicate through
photonic quantum channels.

There exist mainly two methods for entangling distant atoms. One is
based on emission of photons by the first atom, which afterwards
interact with the second atom generating the entanglement
\cite{CirZolKimMab97,EnkCirZol97,GheSaaTorCirZol98,ParKim00,ChaLawEbe02,ChaLawEbe03}.
The second method relies on detecting the photons emitted by the two
atoms with the subsequent entanglement generation due to
interference in the measurement process
\cite{Cabrillo99,BosKniPleVed99}. Some ingredients of both proposals
have been realized experimentally
\cite{Polzik01,Rempe02,Kimble04,Blinov04,Volz05,Kuzmich06,Grangier06,Maunz06,Kimble05,Darquie06}.
Most of the experiments with isolated atoms and light aim at
entangling the internal state of the atom with the polarization of
the photon
\cite{Rempe02,Kimble04,Blinov04,Volz05,Kuzmich06,Grangier06,Maunz06,Darquie06}.
It is clear that due to the size of the Hilbert space, the maximum
attainable entanglement is one ebit.

In this letter we will deal with the generation of entanglement
between two atoms. We will focus on the second method mentioned above,
in which entanglement is generated by measurements. To avoid the limit
of one ebit, we work with continuous variables and seek
entanglement in the motional state of the atoms. We will answer two
fundamental questions: How much entanglement can be produced between
the atoms?  How can we achieve it?

Our first result is that by usual means ---two atoms, one or two
emitted photons, linear optics and postselection
\cite{Cabrillo99,Maunz06}---, we cannot produce more than 1 ebit of
entanglement between the atoms, even if our Hilbert space is larger.
Our second result si that we can achieve an arbitrary amount of
entanglement using at least two emitted photons and what we call an
Entangling Two-Photon Detector (ETPD). The ETPD is a device which
combines both photons in a projective measurement onto a highly
entangled state. Note that this approach differs from recent work on
entangling Gaussian modes of the quantum electromagnetic field by
means of a Kerr medium \cite{Kowalewska06,Olsen06}. Theoretically, an
ETPD could be built using a Kerr medium and postselection, but current
nonlinear materials are too inefficient for such implementation
\footnote{State of the art down-conversion experiments \cite{volz01}
  achieve an efficiency of $5\%$ for weak laser sources.  We expect a
  much lower efficiency for up-conversion due to having much less
  photons and the difficulties of mode matching.}.  Inspired by the
KLM proposal \cite{Knill01}, in the last part of our paper we
demonstrate an efficient scheme for simulating the ETPD using
ancillary photons. Our last result is that introducing $N-2$
additional photons in our setup, together with $N$ single-photon
detectors, beam-splitters and an attenuator, we can obtain an amount
of entanglement of $S=\log_2N$ ebits.  Finally, at the end of the
paper we discuss the relevance of these results and possible
implementations.

We have in mind the setup in Ref.~\cite{Cabrillo99} where two atoms,
initially at zero--momentum state, are excited with a very small
probability. We consider the state of the atoms after spontaneous
emission, when both are in the ground state. The state of the system
at the end is given by
\begin{eqnarray}
  \ket{\Psi} &\sim &
  \epsilon \int dp\, a^{\dagger}_{p} \ket{\text{vac}}( {\cal G}_1(p)\ket{-p,0}+ {\cal G}_2(p)\ket{0,-p}) +
  \nonumber\\
  &+& \epsilon^2 \int dp_1dp_2\, {\cal G}_1(p_1) {\cal G}_2(p_2) a^{\dagger}_{p1}
  a^{\dagger}_{p2}\ket{\text{vac}} \ket{-p_1,-p_2} \nonumber\\
  &+& \ket{\text{vac}}\ket{0,0} + {\cal O}(\epsilon^3)\label{twoatomentini1}.
\end{eqnarray}
Here $p$, $p_1$, $p_2\ldots$ denote the momenta of the emitted
photons; $a_{p,p_1,p_2}^\dagger$ their associated creation operators
and $\ket{\text{vac}}$ the vacuum state of the EM field, and $\epsilon
\ll 1$ are the excitation probabilities of the atoms. The initial
momentum distribution of the emitted photons is given by ${\cal
  G}_i(p)$ for the $i$-th atom. As we will see later, we require some
uncertainty in the initial momentum in order to generate a large
amount of entanglement. Finally $\ket{-p,0}$, $\ket{0,-p}$ and
$\ket{-p_1,-p_2}$ denote the recoil momenta of the atoms after
emitting the photons. The terms omitted in Eq.~(\ref{twoatomentini1})
correspond to higher order processes where an atom emits more than one
photon. These terms will have a very small contribution if the decay
time of the atom is longer than the duration of the exciting pulse.

\begin{figure}
\begin{center}
\includegraphics[width=\linewidth]{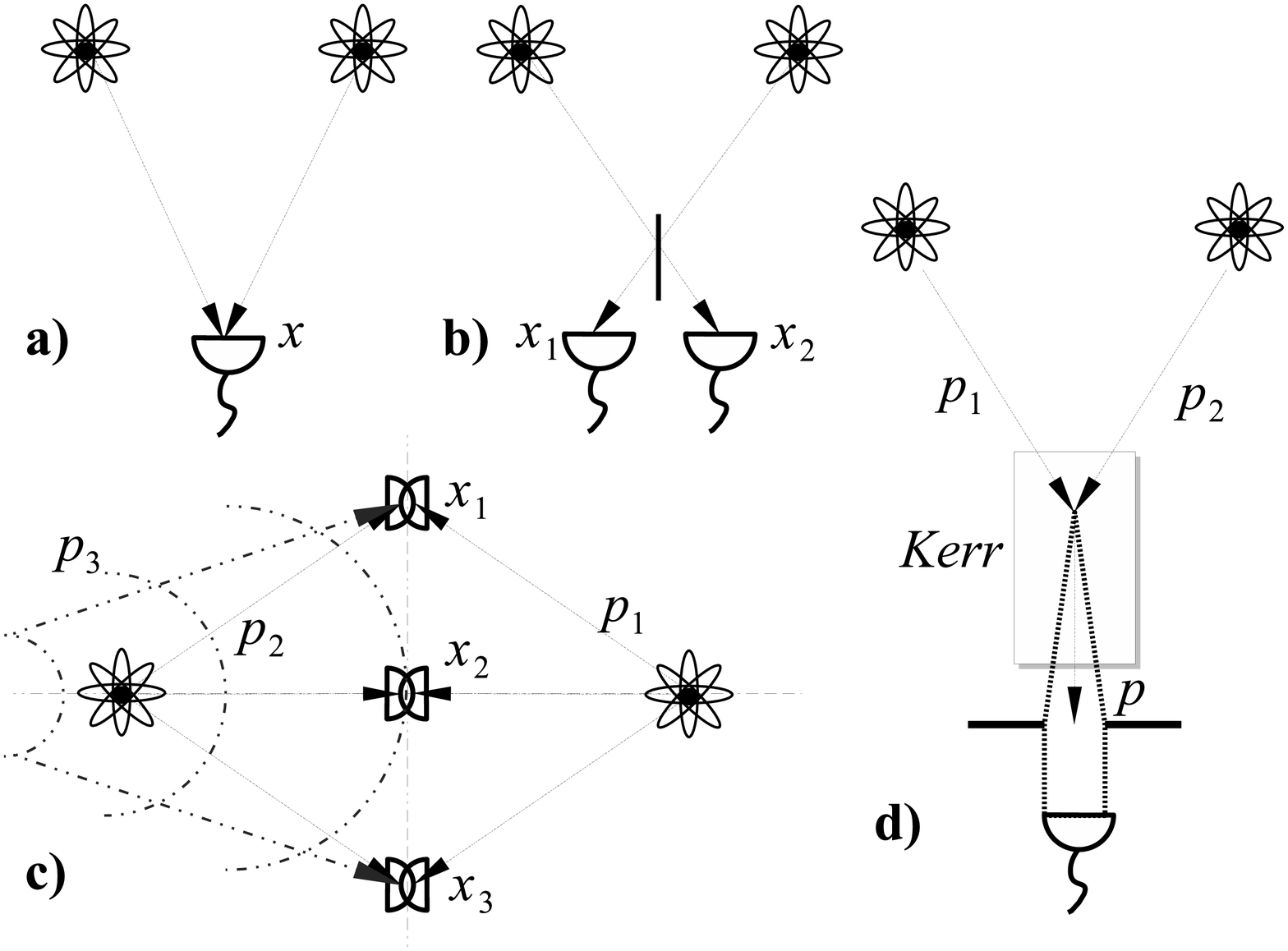}
\end{center}
\caption{Schema of possible experiments for entangling two atoms.
(a)
  Only one photon detected, but we do not know from which atom. (b)
  Two photons are detected, one from each atom. (c) Three photons are
  detected, one being supplied by the experiment (dashed line). Due to
  the setup, the probabilities of reaching each detector are balanced
  and the detectors do not distinguish between left- and right-coming
  photons. (d) Entangling Two-Photon Detector ``gedanken''-experiment.
  By detecting only a range of momenta we entangle the momenta of the
  atoms, $p_{1\perp} + p_{2\perp} \simeq 0$. \label{FigSetup}}
\end{figure}

Let us now consider a single detector placed symmetrically below the
atoms \cite{Cabrillo99}, as in Fig.~\ref{FigSetup}a. If there is one
single photon detection, this will amount to a projective measurement
onto a single-photon state and out of the state in
Eq.~(\ref{twoatomentini1}) only the term on the first row will
survive.  Since the photons coming from the atoms are
indistinguishable, an implicit symmetrization will take place and the
final state of the atoms will be of the form $\ket{\psi_1}\ket{0} +
\ket{0}\ket{\psi_2}$, for some motional states $\psi_1$ and $\psi_2$.
Even though we work with continuous variables, this state can have at
most 1 ebit which corresponds to $\braket{\psi_1 \vert \psi_2} =
\braket{\psi_1 \vert 0}=\braket{\psi_2 \vert 0} = 0$.

We are going to show now that with two emitted photons, linear
optics and two detectors, we cannot do better than one ebit of
entanglement [See Fig.~\ref{FigSetup}b].  The proof generalizes
the previous argument with a little bit more care. First of all,
linear optics amounts to a linear transformation of the initial
momentum modes, $a_p$, to new operators, $b_{\gamma(p)} :=
U_{\gamma} a_p U_{\gamma}^\dagger$. A trivial example of this is a
$50\%$ beam splitter, which changes the photons from incident
states $a_{+p}$ and $a_{-p}$ to $(a_{+p}\pm e^{i\phi}
a_{-p})/\sqrt{2}$.  Linear optics can be combined with
measurements. Without loss of generality, all measurements will
take place at the end of the process and they amount to a
projection onto the modes $a_{x_1}$ and $a_{x_2}$ for the first
and second detector, respectively. The state after a projective
measurement onto two single-photon detectors reads
\begin{eqnarray}
\ket{\Psi_{\rm at}^{(2)}}  & = & \int dp_1dp_2{\cal
G}_1(p_1){\cal G}_2(p_2) \times \nonumber\\ &\times&
\braket{a_{x_1}a_{x_2}b^{\dag}_{\gamma_1(p_1)}b^{\dag}_{\gamma_2(p_2)}}_{\text{vac}}
\ket{-p_1,-p_2}.\label{twoatomenteq2}
\end{eqnarray}
Note that the modes $a_{x_1}$ and $a_{x_2}$ detected by the first and
second detector are expressed on an orthonormal basis different from
that of the $a_p$ or $b$ operators. We enclose this information, plus
the initial wavefunction of the photon in the following $c$-numbers
\begin{equation}
f_j(x_i,p_j) := {\cal G}_j(p_j)[a_{x_i},b^{\dag}_{\gamma_j(p_j)}].\label{twoatomenteq3}
\end{equation}
Using these wavefunctions we define the motional states
\begin{equation}
  \ket{\psi_{ij}} := \int dp\, f_j(x_i,p) |p\rangle.\label{twoatomenteq4}
\end{equation}
The expectation value in Eq.~(\ref{twoatomenteq2}) can be written in
terms of the $f_j(x_i,p_j)$. We thus arrive to the following
expression for the atomic state after the measurement
\begin{equation}
  |\Psi_{\rm at}^{(2)}\rangle \propto
  \ket{\psi_{11}}\ket{\psi_{22}} + \ket{\psi_{21}}\ket{\psi_{12}}.\label{twoatomenteq4bis}
\end{equation}
This state cannot have more than 1 ebit of entanglement, which happens
when all the states $\psi_{11}$, $\psi_{12}$, $\psi_{21}$ and
$\psi_{22}$ are orthogonal to each other.

We must make several remarks. First of all, adding more detectors
does not improve the outcome. Second, our proof is valid
independently of the number of beam splitters, prisms, lenses and
even polarizers we use. In particular, attenuating elements such
as polarizers and filters can be treated as a linear operation
plus a measurement and are covered by the previous formalism.

We propose now to use an ETPD to obtain an arbitrary degree of
entanglement between the two atoms. An ETPD is {\it by definition} a
device that clicks whenever two photons arrive simultaneously and with
their momenta satisfying a certain constraint. An example would be a
parametric up-conversion crystal, in which pairs of photons with
momenta $p_1$ and $p_2$ are converted with a certain probability into
a new photon with momentum $p=p_1+p_2$. One imposes a constraint on
the initial state by post-selecting a window of final momenta.  For
example, restricting the measurement to photons with transverse
momentum $p_{\bot}=0$, then the initial contributing momenta must be
those satisfying $p_{\bot 1}+p_{\bot 2}=0$ [Fig.~\ref{FigSetup}d]. In
this example the ETPD ideally projects the initial two photon product
state $|\Psi^0_{\rm ph}\rangle \!\!= \!\!\!  \int dp_1dp_2 {\cal
  G}_1(p_1){\cal G}_2(p_2) a^{\dag}_{p_1}a^{\dag}_{p_2}|{\rm
  vac}\rangle$ onto the probably entangled state
\begin{equation}
|\Psi^{\rm ETPD}_{\rm ph}\rangle \!\!= \!\!\! \int dp_a dp_b
g(p_a,p_b) a^{\dag}_{p_a}a^{\dag}_{p_b}|{\rm
vac}\rangle.\label{twoatomenteq5}
\end{equation}
Here $g(p_a,p_b)$ is the acceptance function of the detector or,
equivalently, the constraint that the final detected momenta $p_a$ and
$p_b$ obey.

We claim now that with two emitted photons, linear operations and an
ETPD, there is no limit to the attainable entanglement. To prove it we
consider that after projection of the photon part of state in
Eq.~(\ref{twoatomentini1}) into $|\Psi^{\rm ETPD}_{\rm ph}\rangle
\!\!= \!\!\! \int dp_a dp_b g(p_a,p_b)
a^{\dag}_{p_a}a^{\dag}_{p_b}|{\rm vac}\rangle$, the resulting atomic
state will take the form
\begin{equation}
|\Psi_{\rm at}^{\rm ETPD}\rangle = \int
dp_adp_b\,g(p_a,p_b) \ket{\Psi(p_a,p_b)}
\end{equation}
with the already entangled state
\begin{eqnarray}
|\Psi(p_a,p_b)\rangle &:=& \int dp_1dp_2\,[f_1(p_a,p_1)f_2(p_b,p_2)+\\
&&\quad + f_1(p_b,p_1)f_2(p_a,p_2)]|-p_1,-p_2\rangle.\nonumber
\end{eqnarray}
Depending on the specific shape of the functions $g(p_a,p_b)$ and
$f_i(p_l,p_i)$ $l=a,b$, $i=1,2$, the corresponding state may reach an
unbounded degree of entanglement. For example, let us consider that
the photons evolve freely in space without any linear optics elements,
$f_i(p,p_i)={\cal G}_i(p_i)\delta(p-p_i)$, and assume that the
detector has a very narrow acceptance function
$g(p_a,p_b)=\delta(p_a+p_b)$. The wider the initial momentum widths of
the two photons, the larger the resulting bipartite atomic
entanglement, because a higher uncertainty in the wave packets allows
for higher nonlocal correlations, which are not bounded from above.
Indeed, in this ideal case the outcome will be much like the EPR pairs
from the seminal paper Ref.~\cite{Einstein35}.

Current Kerr media are too inefficient to practically implement the
ETPD introduced here. Motivated by this we have designed another
protocol that simulates the outcome of an ETPD using linear optics,
additional photons and postselection. As shown in the KLM proposal
\cite{Knill01}, any highly entangling quantum gate can be performed
this way, though a lot of care is needed to reduce the number of
gates.  Our proposal starts up from the two atoms after having emitted
two photons which are combined with $N-2$ additional ancillary photons,
\begin{eqnarray}
|\Psi^0\rangle & = & \int dp_1dp_2...dp_N
{\cal G}_1(p_1){\cal G}_2(p_2)...{\cal G}_N(p_N)\nonumber\\
&&\times a^{\dag}_{p_1}a^{\dag}_{p_2}...a^{\dag}_{p_N}|{\rm
vac}\rangle\otimes|-p_1,-p_2\rangle.\label{twoatomenteq7}
\end{eqnarray}
The resulting state after linear operations on the $N$ photons,
and $N$-fold coincidence count on the $N$ detectors, will be,
analogously to the two-photon and two-detector case [Eqs.
(\ref{twoatomenteq2})-(\ref{twoatomenteq4bis})]
\begin{eqnarray}
  |\Psi_{\rm at}^{(N)}\rangle=
  \sum_{(i_1,...,i_N)\in\Pi_N}\int
  dp_1...dp_N \prod_k  f_k(x_{i_k},p_k) \times &&\nonumber\\
  \times |-p_1,-p_2\rangle,\label{twoatomenteq8} &&
\end{eqnarray}
where $\Pi_N$ denotes the set of permutations of $N$ elements.  This
state may contain much more than one ebit of entanglement. In fact, an
upper bound to the degree of attainable entanglement is $S=\log_2 N$
ebits. We will show afterwards that this bound is indeed saturated.

\begin{figure}[t]
\begin{center}
\includegraphics[width=0.8\linewidth]{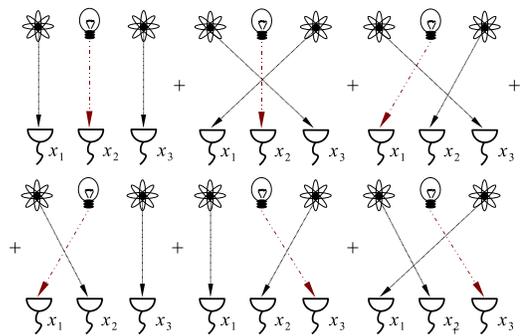}
\end{center}
\caption{Outcome for an experiment with two atoms and three photons,
  as shown in Eq.~(\ref{tresfotones}).\label{Fig1TwoAtoms}}
\end{figure}

As a clarifying example we consider the setup in Fig.~\ref{FigSetup}c
with three photons and three detectors.  Photons $P1$ and $P_2$ come
from their respective atoms, we introduce a single auxiliary photon,
$P_3$ and we place three detectors symmetrically to the atoms, $X_1$,
$X_2$, $X_3$. The final state for the two atoms, considering that all
the three detectors are excited by the three photons, and fixing
relative phases equal to 1 for simplicity purposes, will be
\begin{eqnarray}
|\Psi_{\rm at}^{(3)}\rangle & = & \frac{1}{\sqrt{6}}
(|1,2\rangle+|2,3\rangle+
|3,1\rangle\nonumber\\&&+|1,3\rangle+|3,2\rangle+|2,1\rangle),\label{tresfotones}
\end{eqnarray}
where we denote with $|i,j\rangle$ the atomic state associated to
detection of $P_1$ in $X_i$, and $P_2$ in $X_j$.  In
Fig.~\ref{Fig1TwoAtoms} we show the $N!=6$ processes that contribute
coherently to the two-atom final entangled state. This procedure gives
an entanglement of $S=1.25$ ebits.

\begin{figure}
\begin{center}
\includegraphics[width=7cm]{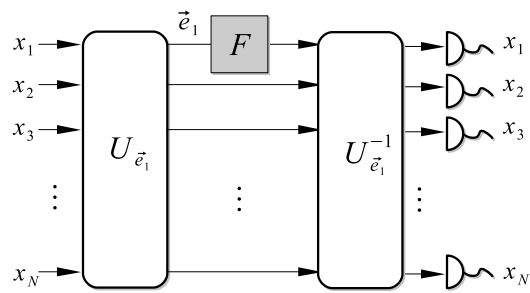}
\end{center}
\caption{Quantum circuit for saturating the bound of $\log_2N$
ebits as described in the text.\label{Fig2TwoAtoms}}
\end{figure}

The previous example is suboptimal. The maximal amount of entanglement
of $S=\log_2 N$ ebits is reachable for some of the states in
Eq.~(\ref{twoatomenteq8}). To prove it we consider a very symmetric
configuration in which the detectors are located along a circle,
equidistant to both atoms [Fig.~\ref{FigSetup}c]. We will assume for
simplicity that the two emitted photons and the $N-2$ ancillary ones
are in $s$-wave states and arrive with equal probability and phase to
every detector. In a similar fashion as in Eq.~(\ref{tresfotones}),
the final bipartite atomic state will take the form
\begin{equation}
\ket{\Psi_{\rm at}^{\rm sym}}= \sum_{ij}C_{ij}\ket{i,j} \propto
\sum_{ij} (1 - \delta_{ij})\ket{i,j},
\end{equation}
where $|i,j\rangle$ is the final bipartite atomic state after
detection of photon $P_1$ in detector $X_i$, and photon $P_2$ in
detector $X_j$. In matrix form, the coefficients $C_{ij}$ are
\begin{equation}
C_{ij}\propto N\vec{e}_1(\vec{e}_1)^T-\openone_{N\times
N},\label{twoatomenteq9}
\end{equation}
where $\vec{e}_1^T:=1/\sqrt{N}(1,1,...,1)_N$. Both the reduced density
matrix of one atom and the Schmidt rank can be obtained from this
matrix. The previous state can be rewritten in the form
\begin{equation}
C_{ij}\propto
(N-1)\vec{e}_1(\vec{e}_1)^T-\sum_{i=2}^N\vec{e}_i(\vec{e}_i)^T,\label{twoatomenteq10}
\end{equation}
where $\{\vec{e}_i\}$, $i=2,...,N$ is a completion of $\vec{e}_1$ to
an orthonormal basis in $\mathbf{C}^N$. From here it is obvious that
the density matrix has full-rank and we can with local operations
obtain a maximally entangled state of the form, up to local phases,
$C_{ij}\propto \openone_{N\times N}$. To do so we must reduce the
contribution of the term $\vec{e}_1$. As shown in Ref.~\cite{Reck94},
a network of beam-splitters and phase-shifters can be used to perform
a unitary operation, $U_{\vec{e}_1}$, that maps the mode
$a_{\vec{e}_1}\propto \sum_{i=1}^{N} a_{x_i}$ to a single optical
port. If, as shown in Fig.~\ref{Fig2TwoAtoms} we place on that port a
filter $F$ that decreases its amplitude by a factor $N-1$, when the
$N$ detectors click simultaneously the atoms will get projected onto a
maximally entangled state with $C_{ij} =
\vec{e}_1(\vec{e}_1)^T-\sum_{i=2}^N\vec{e}_i(\vec{e}_i)^T.$ The proof
is cumbersome and involves studying how all the photon modes in
Eq.~(\ref{twoatomenteq7}) transform under the nonunitary operation
given by the network in Fig.~\ref{Fig2TwoAtoms} and then ensuring that
the detection of $N$ photons does indeed give rise to the maximally
entangled state.

Summing up, in this paper we have demonstrated that it is possible to
achieve an arbitrary amount of entanglement in the motional state of
two atoms by using spontaneous emitted photons, linear optics and
projective measurements. The resulting states can be used to study
violation of Bell inequalities and also as a resource for quantum
information processing. We expect that similar ideas can be used to
entangle atomic clouds, replacing the photons with atoms, because in
this case it is easy to build a two-atom detector.

Regarding the implementation, the ideas shown here can be tested
easily in current experiments. We would suggest using two trapped
ions as target atoms. The ions should be either on a very weak trap,
or released right before excitation. The entanglement in the
momentum will translate into an entanglement in the position of the
atoms after a short time of flight. In practice, with only one
additional photon, $1.58$ ebits can be produced, and we expect a
value of $2$ ebits to be both experimentally achievable and
realistic.  At the cost of a slightly lower fidelity, one can use
$N$ independent attenuated coherent beams instead of true
single-photon sources.  Clearly, even though there is not a
fundamental limit, both the requirement of having good single-photon
sources and the detector efficiency will make it very difficult to
scale this last scheme to larger $N$ and more ebits.

L.L. acknowledges hospitality at the Max-Planck Institute for
Quantum Optics and support from Spanish MEC FPU grant
No.~AP2003-0014 and project No.~FIS2005-05304. J.I.C. acknowledges
support from EU projects SCALA and DFG-Forschungsgruppe 635.

\end{document}